\documentclass[twocolumn,procedia]{article}

\usepackage{graphicx}
\usepackage{titlesec}
\usepackage{amssymb}
\usepackage{amsmath}
\usepackage{multicol}
\usepackage[british]{babel}

\usepackage[figuresright]{rotating}
\usepackage{bm}
\usepackage{caption}

\usepackage{fancyhdr}

\fancyheadoffset[RO,EL]{0pt}
\usepackage{geometry}

\usepackage{siunitx}
\DeclareSIUnit[]\bit{b}
\DeclareSIUnit[]\byte{B}
\usepackage[]{csquotes}

\usepackage{xargs}                      
\usepackage[pdftex,dvipsnames]{xcolor}  
\usepackage[colorinlistoftodos,prependcaption,textsize=tiny]{todonotes}

\usepackage{url}

\usepackage[backend=biber,sorting=none]{biblatex}

\newcommandx{\unsure}[2][1=]{\todo[linecolor=red,backgroundcolor=red!25,bordercolor=red,#1]{#2}}
\newcommandx{\change}[2][1=]{\todo[linecolor=blue,backgroundcolor=blue!25,bordercolor=blue,#1]{#2}}
\newcommandx{\info}[2][1=]{\todo[linecolor=OliveGreen,backgroundcolor=OliveGreen!25,bordercolor=OliveGreen,#1]{#2}}
\newcommandx{\improvement}[2][1=]{\todo[linecolor=Plum,backgroundcolor=Plum!25,bordercolor=Plum,#1]{#2}}
\newcommandx{\thiswillnotshow}[2][1=]{\todo[disable,#1]{#2}}

\titleformat*{\section}{\normalsize\bfseries}
\titleformat*{\subsection}{\normalsize\itshape}

\begin{document}

\title{
\begin{flushleft}
A novel approach for FPGA-to-server data transmission over an Ethernet-based network using the eXpress Data Path technology
\end{flushleft}
}
\date{}

\author{Carsten Dülsen, Tobias Flick, Timo Göhring, Wolfgang Wagner, Marius Wensing \\
  {\itshape\footnotesize University of Wuppertal, Gau{\ss}stra{\ss}e 20, 42119 Wuppertal, Germany}}





\maketitle

\begin{abstract}
\small
In the context of the upgrade of the Large Hadron Collider at CERN for high-luminosity operation, the particle detectors have to cope with much higher data rates and therefore need to upgrade their data acquisition systems.
This upgrade is taken as an opportunity to exchange the currently used highly customized hardware by commercial solutions.
Nevertheless, some part of the data processing still needs to be done within Field Programmable Gate Arrays (FPGA), requiring the transfer of data between the FPGAs and the commercial servers.
This paper reports on a study of direct data transmission from FPGAs to servers via a commercial network. 
Large data buffers as required for reliable data-transmission protocols are avoided by using an emerging technique named eXpress Data Path (XDP).
Based on XDP, the transmission of \SI{5,2}{\peta\byte} (i.e. \SI{2,92e12}{packets}) was achieved within \SI{168}{\hour} without a single missing packet.
\end{abstract}

\section{Introduction}

The Large Hadron Collider (LHC) at CERN is the world's most powerful particle accelerator, searching for rare or even unknown particles and performing precision measurements of standard model processes.
To improve the capability of detecting even more rare particle processes, the LHC will be upgraded within the next decade, resulting in an increased number of concurrent proton-proton collisions, while keeping the time between proton bunch crossings at \SI{25}{\nano\second}, i.e. an event rate of \SI{40}{\mega\hertz}.
As a consequence, the experiments ATLAS and CMS will be upgraded in two steps (i.e. \emph{Phase I} and \emph{Phase II}) to cope with the higher number of particles produced in one collision event.
An important part of the Phase II detector upgrades is the replacement of the inner tracking systems.
The spatial granularity of the sensors is increased to resolve the hits produced by the high number of charged particles.
The increase in track density and detector granularity leads to more data being generated for the individual collision events.
At the same time, the trigger rate is increased from \SI{100}{\kilo\hertz} to \SI{1}{\mega\hertz} for ATLAS and \SI{750}{\kilo\hertz} for CMS~\cite{ATLAS:TDAQPhaseIITDR,CMS-TDR-022}.
The data acquisition (DAQ) systems of ATLAS and CMS will be upgraded to cope with these challenges.

Due to the geometry of particle detectors and the requirement of a low mass implementation, only a limited set of cables with a fixed bandwidth can be used.
The same limitations apply to the amount of memory being incorporated into the data acquisition (DAQ) front-end electronics of the detector system.
These boundary conditions lead to the requirement of data transmission without any capability for resending data.
The loss of data packets must be as low as possible.

Communication with the detector is subject to constraints such as precise timing, 
radiation tolerance of the front-end electronics, a low material budget to keep multiple scattering at a low level, 
limited bandwidth, and error tolerance, requiring the usage of custom protocols.
Therefore, the DAQ systems make use of Field Programmable Gate Arrays (FPGAs) in the first DAQ stage for being able to process these protocols while keeping the flexibility of a programmable system.
All subsequent stages can be implemented in software.

\begin{figure*}[t]
  \centering
  \includegraphics[width=.97\linewidth]{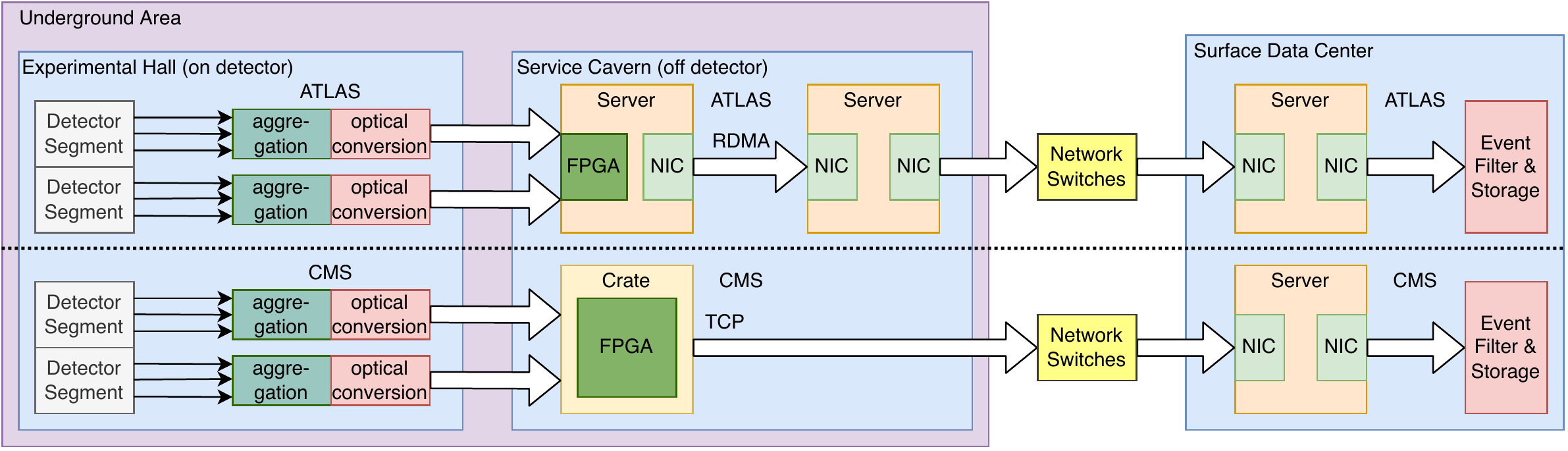}
  \caption{The two designs of the readout and process chains used by the ALTAS and CMS
    collaborations~\cite{ATLAS:TDAQPhaseIITDR,CMS-TDR-022}. The event data from the detector elements goes through several processing steps before it is written to the long term storage. The ATLAS experiment makes use of PCIe cards within commodity servers (top chain) for the protocol translation, while CMS favors the usage of a MicroTCA system (bottom chain).}
  \label{fig:ReadoutChain}
\end{figure*}

Once data is received by the FPGA it has to be transmitted to a server farm for further processing and event reconstruction implemented in software.
This paper describes a proposal for the data transmission from the FPGA-cards to servers meeting the constraints of the future Inner Tracker (ITk) of the ATLAS detector~\cite{ATLAS:PixelTDR}.

\section{Motivation}
With the high-luminosity detector upgrades the DAQ chain will be renewed to accommodate the much higher data rates of \SI{5.2}{\tera\byte\per\second} for ATLAS and \SI{51}{\tera\bit\per\second} for CMS~\cite{ATLAS:TDAQPhaseIITDR,CMS-TDR-022}.
In the past years, the LHC experiments decided to move from custom designs for the off-detector part 
of their DAQ systems to commercially available (commercial off-the-shelf, COTS) components, especially for the communication part.
The only custom-made part in the off-detector sections of the DAQ chains is an FPGA card converting between detector specific and industry-standard protocols, and performing further aggregation steps to higher-bandwidth links to facilitate the data processing
in commercial networks and PCs.
All further processing is done by software on commodity servers.
The designs of ATLAS and CMS are depicted in Figure~\ref{fig:ReadoutChain}.


These requirements place certain constraints on the design of the DAQ systems.
The output bandwidth of the FPGA has to be higher than its input bandwidth in order to not block the data transmission.
Also, space and power is limited in the service cavern and therefore, a small-sized system is highly desirable.
Another requirement of the DAQ system is its reliability in terms of data transmission.
Over the lifetime of the detector data losses should be less than \SI{1}{\percent} in the transmission
and storage of data, including losses due to radiation inside the detector~\cite{ATLAS:PixelTDR}.

\section{Upgrades of the DAQ for Phase I + II}

For the Phase I + II upgrades, the ATLAS and CMS experiments have chosen different methods to transfer the event data from the FPGAs in the service caverns to commodity servers for further processing.
This transmission step is depicted in Figure~\ref{fig:ReadoutChain} as the part within the service cavern.

\subsection{The ATLAS DAQ upgrades}
For the Phase I upgrade, the ATLAS experiment decided to use a 16-lane third-generation PCI-Express (PCIe) link to transfer the data from the FPGA cards into the host memory and further over the commercial network to the servers for processing~\cite{Gottardo:2746729}.
For the Phase II upgrade, the system will be replaced by an updated version.
The discussion about the PCIe standard to be used (PCIe generation 5 is favored) is still ongoing, but the data transmission scheme will stay the same.

In the development of the Phase I upgrade, the protocol used for transmission was switched from TCP to Remote Direct Memory Access (RDMA) since the processing of the TCP protocol generates a large CPU workload and data is copied several times.
The RDMA protocol offers the reception of data without any involvement of the CPU because the network interface card (NIC) stores the data directly into a buffer being prepared by the receiving software within the host memory of the receiving server.
For the Phase II upgrade, research work has started in ATLAS to study the usage of RDMA inside the FPGA. Promising
first studies have been performed, getting close to the theoretical bandwidth limit of the communication
link~\cite{Vasile:2023qch}.

While the usage of RDMA helps to reduce the workload of data reception, the DAQ system still has to store the data on the sending side until the correct transmission is acknowledged.

\subsection{The CMS DAQ upgrades}
The CMS experiment uses a MicroTCA system holding the FPGAs for the Phase I upgrade.
The network interface is implemented in the firmware using TCP.
TCP was developed as an internet protocol and thus covers the complete path from the sender to the recipient, including all its subsections.
The available bandwidth for a transmission needs to be constantly probed since the available bandwidth for each subsection is not known in advance and can vary over time.
This is done by increasing the sending rate until the fraction of dropped packets is getting too high, making retransmission a key requirement of the protocol.
Furthermore, the algorithms used by the TCP protocol are optimized for a software implementation.
Therefore, CMS developed a simplified version of TCP on the sending side,
but is still compatible with the standard on the receiving side, such that a commodity server
running an unmodified Linux can be used.
As in the case of RDMA the usage of TCP also requires sufficient memory in the sending FPGA.

The current system uses a single \SI{40}{\giga\bit\per\second} port consisting of four \SI{10}{\giga\bit\per\second} links per FPGA.
The FPGA provides enough internal memory resources to store the data for possible retransmissions, so no external memory is needed.
For the Phase II upgrade~\cite{CMS-TDR-022}, these \SI{40}{\giga\bit\per\second} ports are planned to be replaced by several \SI{100}{\giga\bit\per\second} ports (4*\SI{25}{\giga\bit\per\second} each) per FPGA.
The Xilinx VU35P FPGA is chosen since it features sufficient built-in high bandwidth memory (HBM) required for
buffering data to ensure a lossless transmission from the off-detector cards to the event builder.

\subsection{Motivation for studies without buffering of data}
The described usage of the TCP and RDMA protocols in the DAQ of ATLAS and CMS is different from
what the protocols were originally developed for. In this sense, they may not the most ideal choice.  
Both solutions require the buffering of data until the receiving side acknowledges the reception of the
data. 
The memory needed for the buffering requires the use of specific FPGAs with a large amount of memory,
or the design of a complex interface to high-speed external memory.
In addition, TCP constantly probes the available bandwidth of the transmission chain. 
However, for the LHC experiments, the bandwidth of the optical data links between the front-end electronics and the
off-detector hardware defines the maximum data rate of each stream.
Exploiting this knowledge, the subsequent network can be specifically optimized to have enough bandwidth for all transfers. 
Furthermore, reliable data transmission is not guaranteed over the full data path at the LHC experiments,
since the detectors are not able to deliver error free data all the time. 
The reasons for this are the radiation environment and general operational issues.
Nevertheless, each component should contribute as little as possible to the data loss.
However, the inevitable occurrence of data loss poses the question whether the implementation of
reliable data transmission between the FPGA and the higher level DAQ servers is necessary or
whether this requirement can be lifted, resulting in a considerable reduction of system costs with
a sufficiently small amount of data being lost.
Choosing an unreliable data transmission scheme suggests the usage of UDP which in addition has the benefit of requiring significantly less processing power.
 
\section{Data reception in Linux systems}
Because the Linux network stack is optimized in terms of functionality and performance for average usage, it lacks performance in more specific scenarios.
Most of these scenarios have to handle either a very high packet rate or very high data throughput.

A way to increase the performance is to bypass the Linux kernel and its network stack and to handle the whole chain in userspace.
This means, that the user programs take control of all involved components including the network card.
Such a solution requires the user programs to also include the corresponding drivers.
Frameworks like the Data Plane Development Kit (DPDK) thereby provide an abstraction layer to abstract also from the operation system~\cite{DPDK}.
The downside of such solutions is that the user program needs to handle the complete network traffic, including network traffic not meant for the specific application.
This traffic is simply dropped since the re-injection into the Linux network stack would generate too much overhead.
Therefore, such solutions are only viable if the framework has access to a separate NIC.

\begin{figure*}[!t]
  \centering
  \includegraphics[width=1\linewidth]{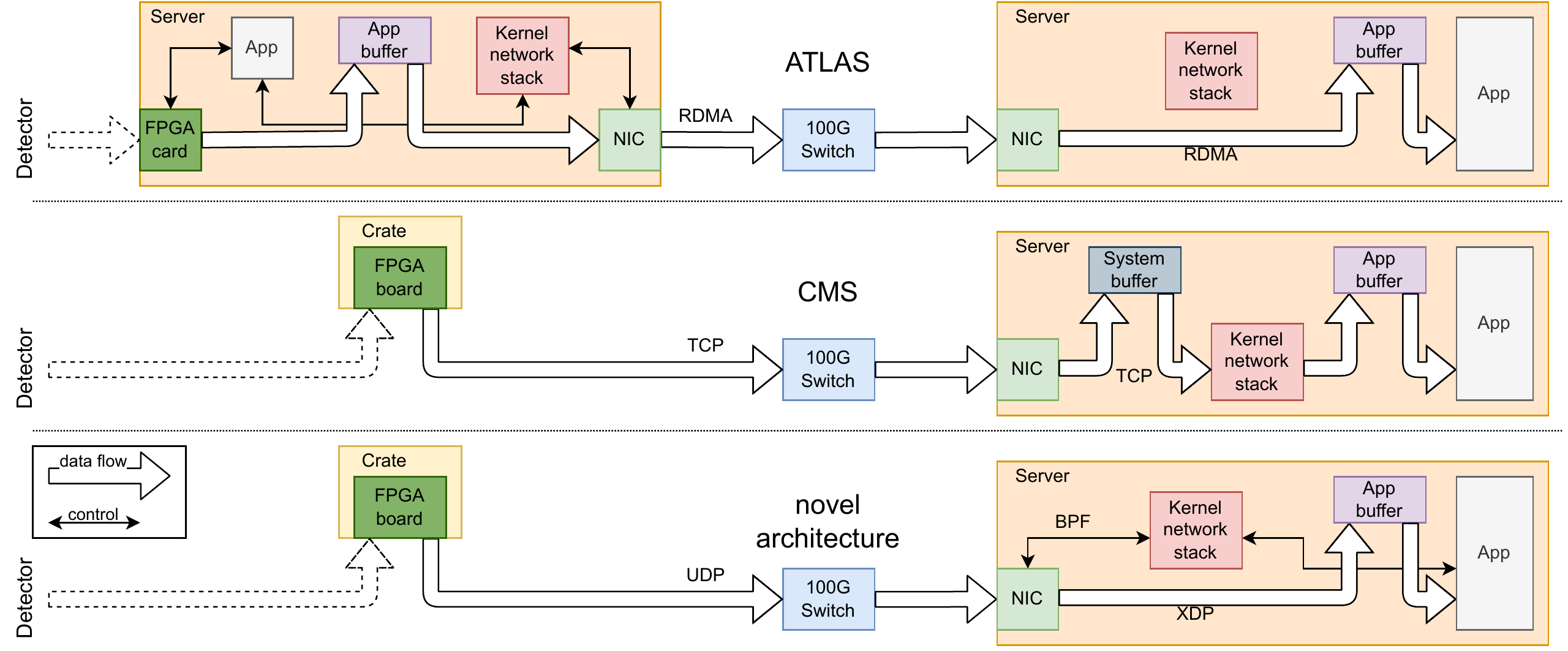}
  \caption{Comparison of the DAQ architectures used by the ATLAS experiment (top) and the CMS experiment (center) versus the novel architecture proposed by this paper (bottom). While the architecture used by ATLAS favors simplicity on the receiving side, the CMS experiment focused on the optimization of the sending side. The novel architecture combines both approaches and applies additional optimizations to reduce the needed resources even further.}
  \label{fig:OverviewArchitecture}
\end{figure*}
To solve this issue, a novel technique called \emph{eXpress Data Path (XDP)}~\cite{10.1145/3281411.3281443} was developed 
for the Linux kernel.
This technique offers a fast lane which bypasses the Linux network stack and reduces the processing overhead for data reception of selected network traffic.
The remaining traffic is still handled by the Linux network stack.
XDP makes use of a virtual machine called Berkeley Packet Filter (BPF)~\cite{BPF} within the Linux kernel to filter the arriving network traffic right after the NIC issues the network interrupt.
This filtering is based on the receive queues of the NIC and results in one of the predefined actions:
\begin{itemize}
	\item drop the packet
	\item send it out on the same NIC again
	\item redirect to another interface
	\item redirect to an XDP socket
	\item forward it to the network stack
\end{itemize}
In case of redirection to an XDP socket, the selected packets are moved into a memory area prepared by the user application.
The communication via the socket is therefore reduced to manage the individual packet slots within this memory area, for example notifications of filled slots or marking them as available again.
The drawback of this technique is, that no protocol decoding or error checking is done by the Linux network stack and therefore these tasks need to be implemented in the user application.
This new technique is mainly used for network management or filtering software or to accelerate the network interfaces of virtual machines.
No prior case of its use for improving direct data reception is known to the authors of this paper.

\section{Novel architecture}
To overcome the limitations of the current DAQ systems, a novel architecture is proposed in this paper which is shown in Figure~\ref{fig:OverviewArchitecture}.
The data are transmitted directly from the FPGA into the network as in the CMS experiment, but the UDP protocol is used rather than the TCP protocol.
The direct transmission enables a better scalability of the system compared to the usage of a PCIe card with a fixed bandwidth, while usage of the UDP protocol reduces the system complexity by obviating the need for additional memory components due to the omission of guaranteed data transmission.
The combination of both optimizations results in a reduction of the required resources (space, power, costs).

However, the simplicity of the UDP protocol comes at a price:
In UDP, each packet stands for itself and there is no information about the order of packets or completeness included in the protocol.
This also means that the information of the UDP protocol header is not enough to determine the fraction of lost packets.
To be able to determine this fraction, additional information is needed to establish a fixed relationship between the network packets.
The easiest solution for this problem is the usage of a packet identifier field which is filled by a packet counter\footnote{Also a pseudo random ordering can be used as long as the receiver can determine the expected counter value.}.
In contrast to the identifier of the IP protocol which is increased by all packets of the same network interface, the identifier implemented here is used for each individual data stream and therefore the stream processors should see continuously increasing values.
This packet identifier is implemented in a new protocol called Data ReadOut Protocol (DROP)~\cite{Dulsen:2779997}.
A retransmission of lost packets is deliberately not included in this protocol to avoid 
a requirement for additional memory in the FPGA.

\begin{figure}[tb]
  \centering
  \includegraphics[width=1.0\linewidth]{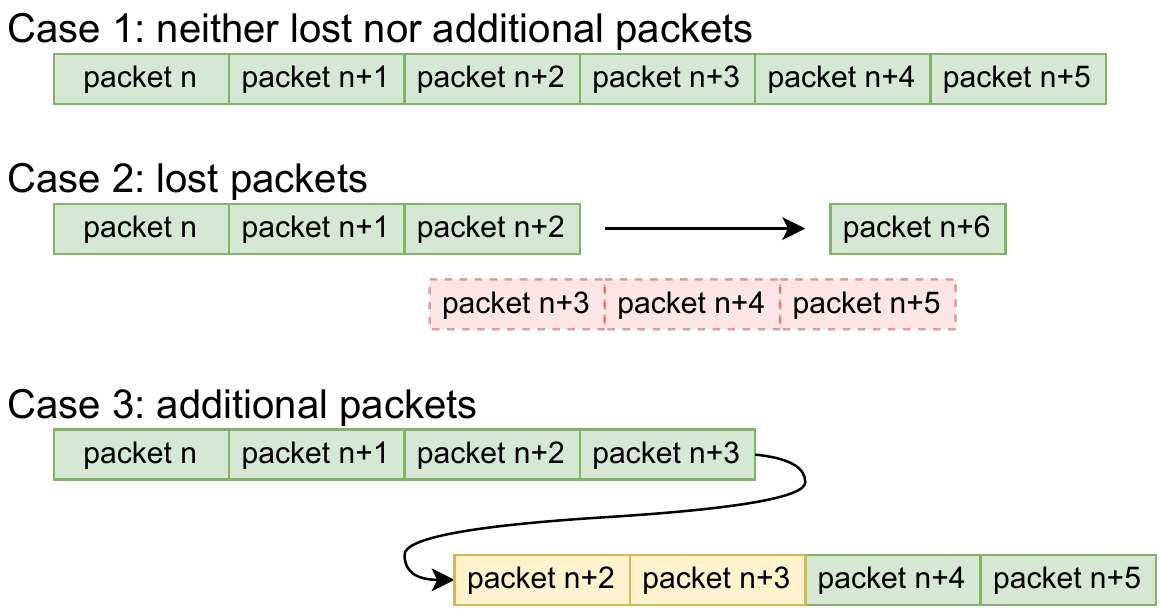}
  \caption{The packet identification field can be used to validate that each packet was received exactly once and in the correct order. All deviations from this can be described by a set of jumps in the value of the identifier.}
  \label{fig:DropStream}
\end{figure}

Even if the content of subsequent packets is independent of each other and therefore no ordering is required, the ordered structure of the identifier field eases the check of packet loss as depicted in Figure~\ref{fig:DropStream}.

With the packet identifier being counted up, the packet identifiers of consecutive packets should have a distance of exactly 1.
There are two error cases if only the distance between the packet identifiers of consecutive packets are checked:
\begin{itemize}
	\item Is the distance is larger than 1, the packets in between are regarded as being dropped.
  \item Is the distance 0 or negative, it is assumed that additional packets are seen.
\end{itemize}
More complex patterns of unexpected behavior can be modeled by combinations of the two cases.
For example, the out of order delivery of a packet would result in three error reports:
\begin{enumerate}
	\item a distance larger than 1 is reported (i.e. the delayed packets are reported as missing)
	\item a negative distance is reported (i.e. the delayed packets are now reported as additional packets)
	\item a distance larger than 1 is reported (i.e. the early packets are now reported as missing)
\end{enumerate}
This example shows that certain combinations of such error reports can still indicate a lossless data transmission.
However, the intention is to achieve a data transfer without any of such reports.

On the receiving side, the XDP technique is used to achieve a reduction of the CPU load for data reception similar to the RDMA used by the ATLAS experiment.
XDP redirects the incoming network packets to the software instance dedicated to the data stream as described in the previous section.
To steer the redirection by the BPF program, the received packets are sorted by their UDP destination port into the corresponding receive queues.

\begin{table}[b]
	\caption{Absolute utilisation of the different firmware parts for a \emph{Xilinx} Virtex Ultrascale+ FPGA.}
	\label{tab:utilisation}
	\resizebox{\linewidth}{!}{%
		\begin{tabular}{lrrr}
			& \multicolumn{1}{c}{\textbf{Ethernet sub-}} & \multicolumn{1}{c}{\textbf{UDP/IP}} & \multicolumn{1}{c}{\textbf{DROP}}     \\
			& \multicolumn{1}{c}{\textbf{system (100G)}} & \multicolumn{1}{c}{\textbf{stack}}  & \multicolumn{1}{c}{\textbf{protocol}} \\ \hline
			LUTs             & 10100                                      & 5600                                & 1900                                  \\
			Flipflops        & 14100                                      & 15800                               & 2800                                  \\
			Block RAM (kBit) & 144                                        & 918                                 & 0                                     \\ \hline
	\end{tabular}}
\end{table}

To reach maximum performance of the receiving program, it has to be optimized for the architecture of the used CPUs.
Also, the receiving program needs to be split up into threads to ensure that each task can work as independently as possible and not be blocked by other tasks.
Since the transfer of data between the CPU cache and the host memory needs a considerable amount of time, the whole protocol processing should happen within the CPU caches.
This requires good knowledge of the architecture of the CPU, especially about the connections between the CPU cores and the different cache blocks.
For test setup under study an optimized assignment of tasks to CPU cores is presented
in Section~\ref{sec:optimization}.

\section{Test setup}

\begin{figure}[t]
  \centering
  \includegraphics[width=1.0\linewidth]{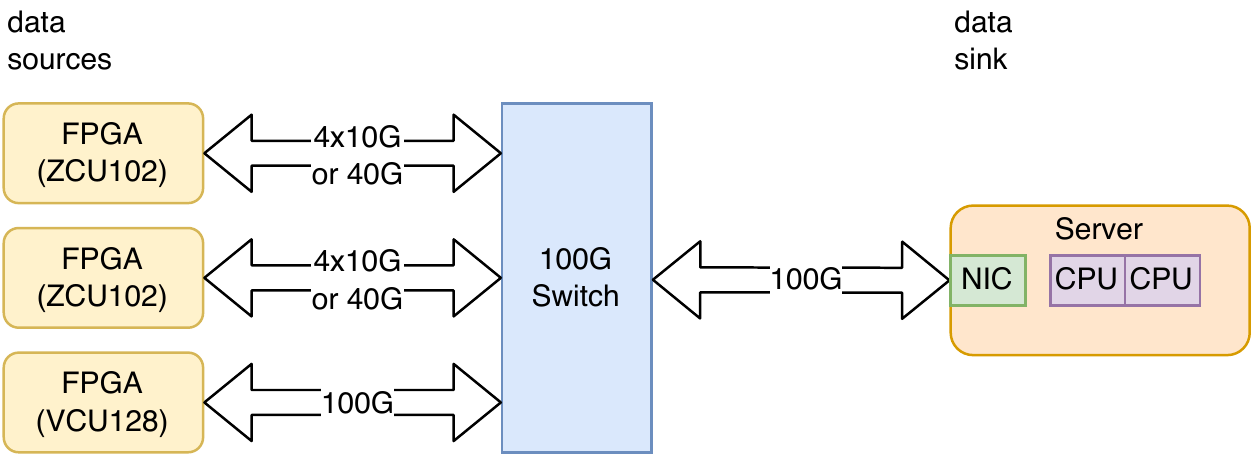}
  \caption{The test setup consists of up to three FPGA evaluation boards with different link speeds as data sources, a \SI{100}{\giga\bit\per\second} Ethernet switch and the commodity server being used as data sink.}
  \label{fig:TestSetup}
\end{figure}

To determine the fraction of lost packets and evaluate the new approach, a test system was set up as shown in Figure~\ref{fig:TestSetup}.
The setup consists of \emph{Xilinx} FPGA evaluation kits as data sources and a commodity server as data sink.
The boards used are the \emph{ZCU102}, featuring a Zynq Ultrascale+ FPGA, and 
the \emph{VCU128} with a Virtex Ultrascale+ FPGA.
Since the \emph{ZCU102} supports only a single \SI{40}{\giga\bit\per\second} Ethernet connection, two of these are used to reach the intended data rate.
This setup has the benefit of having two independent data sources whose data streams needed to be interlaced.
The \emph{VCU128} supports up to four \SI{100}{\giga\bit\per\second} Ethernet connections from which one is used for presented tests.
The firmware of the FPGA consists of an UDP/IP network stack with a line rate of \SI{40}{\giga\bit\per\second} or \SI{100}{\giga\bit\per\second} (i.e. 40GbE and 100GbE) depending on the firmware version, a module implementing the DROP protocol and multiple data generators, producing a data rate of up to \SI{10.24}{\giga\bit\per\second} each~\cite{Dulsen:2779997}.
The resources utilised inside the FPGA for implementing the network stack are given in Table~\ref{tab:utilisation}.
The data generator can be configured using the following parameters:
\begin{itemize}
	\item \emph{block size} specifying the amount of data to be generated,
  \item \emph{packet size} for the maximum size of the parts the block is split into\footnote{The last packet may not reach the full size since the remaining number of bytes within the block is limited.},
  \item the selected \emph{pattern type} (e.g. \SI{16}{\bit} counting pattern).
\end{itemize}
Furthermore, the generated data rate can be reduced by introducing artificial pauses after each data word or at the end of a packet.
The output of the data generators is then packed into DROP streams with the packet identification field being counted up for each packet.
Thereby, the output of several data generators can be packed into the same DROP stream when a data rate higher than \SI{10.24}{\giga\bit\per\second} for a single data stream is needed or the data stream should include packets of different sizes.

The FPGA boards and the servers were connected by a 100GbE network switch.
The network was reserved for the presented tests and not used for other services, and had no direct connection to other networks.
In addition, the server was connected to a \SI{1}{\giga\bit\per\second} management network used to access the server from remote.

The used server is equipped with two \emph{AMD EPYC 7302} CPUs with 16 cores / 32 threads each and a clock frequency of up to \SI{3.3}{\giga\hertz}.
A \mbox{\emph{Mellanox ConnectX-5}} 100GbE NIC was used as network interface.
The operation system is \emph{Fedora} in version 30 and the Linux kernel version is 5.2.9.
To reduce the interruptions by the remaining system, a complete CPU\footnote{The one which provided the PCIe lanes used by the NIC.} of the server is isolated from the Linux scheduler such that no other process than the ones needed for the test is running on these.

\begin{figure}[t]
  \centering
  \includegraphics[width=1.0\linewidth]{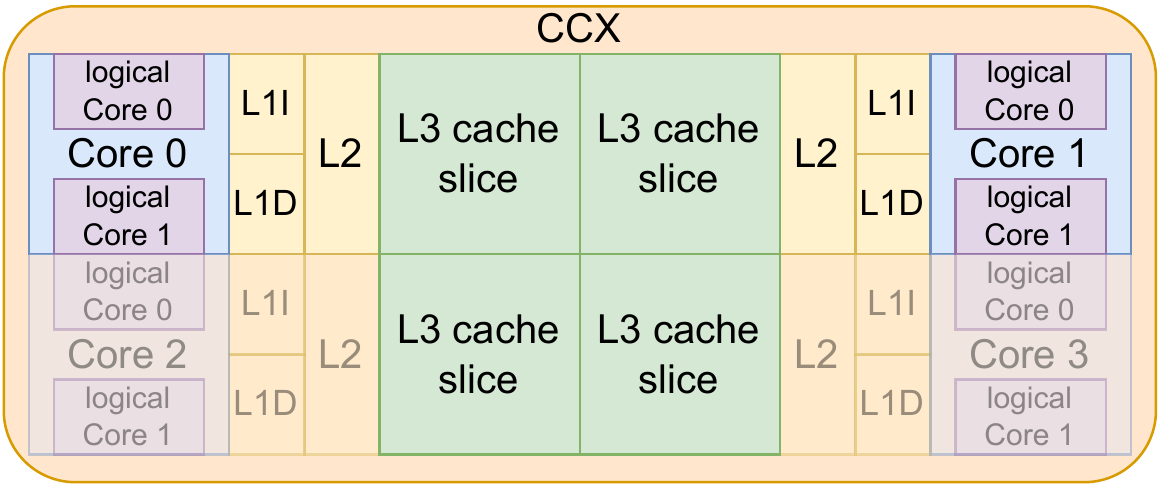}
  \caption{In general, a core complex (CCX) of the AMD Zen2 architecture~\cite{AMD_EPYC_Arch}
    houses four CPU cores and their CPU caches. In the used variant of the CPU, only two out of the four cores are activated, while all four L3 cache blocks are available.}
  \label{fig:AMD_EPYC_CCX}
\end{figure}

The \emph{AMD Zen2} architecture used in the \emph{EPYC} CPUs groups processor cores into
\emph{core complexes (CCXs)}~\cite{AMD_EPYC_Arch} with a combined level 3 cache for all cores within the CCX as shown in Figure~\ref{fig:AMD_EPYC_CCX}.
To avoid unnecessary data transfers all threads of an instance of the user program including the network interrupt running the BPF program were placed on the same CCX.

The user program was split into a thread taking care of the XDP sockets as well as monitoring the incoming data (i.e. checking the packet identifier, counting received packets and bytes) and the threads histogramming the payload (i.e. counting the occurrence of each \SI{16}{\bit} word).
The histogramming is done to include a workload similar to the decoding of the data streams coming from the detector elements.
Since the test data consists of simple counting patterns, there is nothing to be decoded.
However, the histogramming ensures that each byte of the payload is accessed and therefore the whole packet was loaded into the CPU cache.
The management thread is common for all data streams being received by the instance of the user program while the threads histogramming the data serve a single data stream each.
To keep the data locally in the cache of the CCX, an individual instance of the user program is started for each CCX used.
In other words, only a single instance of the user program should run on a CCX to make best usage of its limited resources.

\section{Measurements}

The described setup was used to perform a series of measurements which can be grouped into three phases.
The first phase addresses the total maximum packet rate (measured in million packets per second, \si{\mega p\per\second}) the server can handle.
It is assumed that this packet rate only depends on the number of packet headers (i.e. the number of packets) to be processed.
Since the memory for the packet buffers is pre-allocated, the payload size is assumed to have no impact on the total maximum packet rate as long as a constant data rate is maintained.
With the maximum user data rate being the product of the maximum packet rate and the payload size used, these measurements are important for determining the minimum payload size needed to achieve a given data rate.

The second phase deals with the question of how many CPU cores are needed to process the total data rate.
The processing capability of a single CCX is evaluated by investigating the packet loss for different assignments of the threads.
The result of this group of measurements is a set of rules on how to assign the different threads to reach the best performance as well as a recommendation about the distribution of processing of individual data streams on the used CPU.

In the third phase, the results of the other phases are validated by performing long term measurements.
The long duration is necessary since the limit on the packet loss rate corresponds to the total amount of transferred data.

\section{Measurement: Maximum packet rate}

Since the processing of the protocol headers is expected to be the most time-consuming part of the data reception process, the number of packets (and therefore the number of headers to be processed) which can be received in a given time period is an important characteristic of the server.
This value defines the minimum packet size needed to reach an intended data rate.

The measured packet rates are expected to vary around a fixed point.
Since only the order of magnitude of the fixed points is of interest, the duration of the tests was only a few minutes for each measurement until the results settled.

\subsection*{Measurement: packet rate with packet loss}

To measure the total maximum packet rate of the server, a very large number of packets was sent at a very high rate to the server and the number of packets seen by the receiving software was recorded.
The data generators were configured to send packets with a payload size of \SI{64}{\byte} at the highest possible rate, resulting in a packet rate of around \SI{16}{\mega p\per\second} per data stream~\cite{TG:bachelorthesis}.
Since a single CPU core is expected to be unable to cope with the total packet rate, the number of active data streams (i.e. pairs of data generators and receive programs) was increased during the test while the configuration of the data generators was fixed.
On the receiving side, the number of recognized packets per data stream was measured and the sum over all streams was taken.
The total maximum packet rate is expected to saturate at a constant value for the packet rate, which marks the end of this test.
Since it is obvious that packets will get lost in these measurements, the fraction of lost packets is ignored.

\begin{figure}[t]
  \centering
  \includegraphics[width=1.0\linewidth]{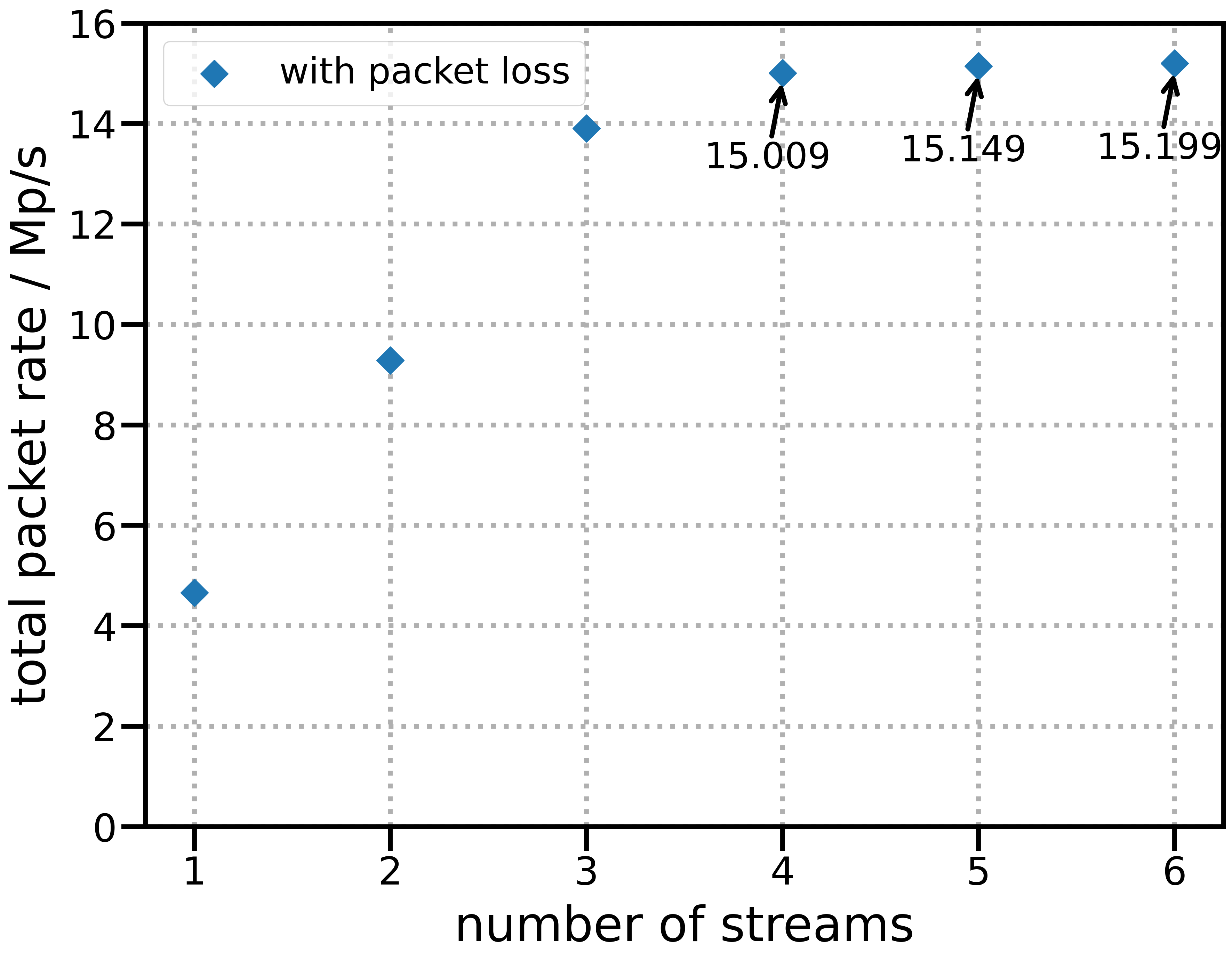}
  \caption{Measured packet rate for different numbers of data streams for packets with a payload size of \SI{64}{\byte}. The results for up to three data streams show a good linearity, i.e. the additional streams do not interfere with the existing ones. Starting with four data streams, the total packet rate saturates and reveals the limits of the system.}
  \label{fig:PacketRate}
\end{figure}

The results of the measurement are shown in Figure~\ref{fig:PacketRate}.
A single data stream achieved a maximum packet rate of \SI{4.66}{\mega p\per\second} without any decoding and \SI{3.88}{\mega p\per\second} when the decoding was turned on.
With the increasing number of data streams used, the corresponding multiple of these values were measured, showing the limitation by the CPU cores.
At a rate of around \SI{15}{\mega p\per\second}, the total packet rate saturated with four data streams and remains quite stable when the number of data streams is further increased.
Therefore, the maximum packet rate the server can handle is determined to be at the level of \SI{15}{\mega p\per\second}.

\subsection*{Measurement: packet rate without packet loss}

Since the first measurement ignored the fraction of lost packets, the measurements were repeated in a slightly different way to get the limit on a lossless transmission.
The packet rate was varied by changing the length of the pause between consecutive packets while the packet size and the number of data streams were kept constant.
The maximum packet rate of each data stream was chosen such that the accumulated packet rate of all streams exceeded the total maximum packet rate of the server measured in the first measurement by about \SI{10}{\percent}.
The payload size was set at a value such that the maximum packet rate per data stream is reached by a generator when only the minimum pause length is applied.
This results in the use of five data streams, a maximum packet rate per stream of \SI{3.4}{\mega p\per\second} and a value of \SI{350}{\byte} for the payload size.
The length of the pause between packets was set to very high values (i.e. low packet rates) at the start of the tests and was reduced for each measurement to increase the total packet rate.
Since the total packet rate of all data streams is set to be larger than the maximum packet rate of the server, packets are expected to start to get lost at some point and the total packet rate saturates at the total maximum packet rate of the server.

\begin{figure}[t]
  \centering
  \includegraphics[width=1.0\linewidth]{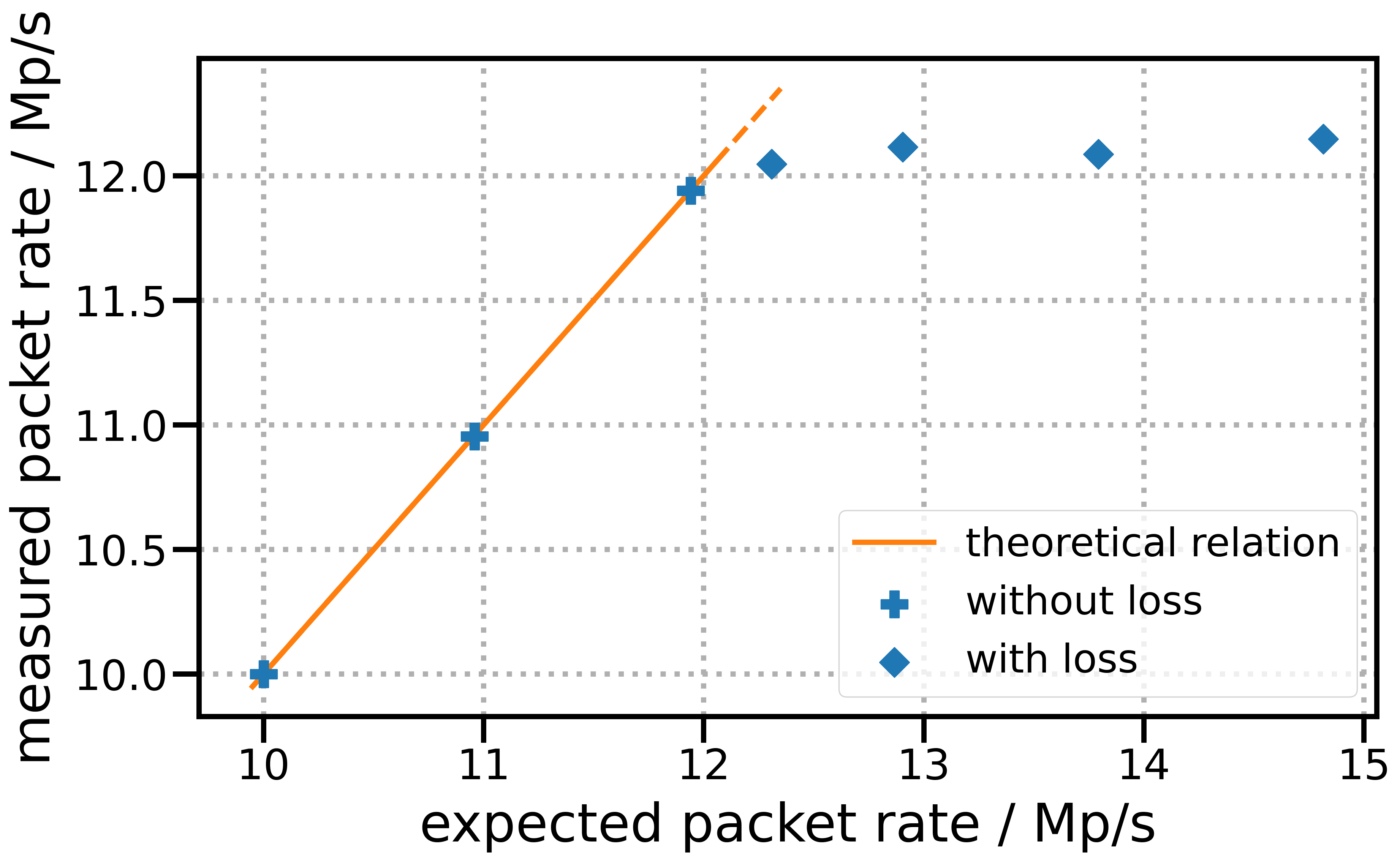}
  \caption{The relation between measured and expected total packet rate for the repeated measurements with five data streams, a payload size of \SI{350}{\byte} and with an artificial delay of different length. The length of the delay was reduced for each individual measurement until the ratio between both differs from one. The limit on the total packet rate was already found at a lower point with respect to the measurement with a payload size of \SI{64}{\byte}. At that point, length of the delay has not reached zero meaning that the limit is not caused by the sending side i.e. the full data rate of the data generators is not reached.}
  \label{fig:PacketRate2}
\end{figure}

In this second measurement, the data was transferred without packets being dropped for packet rates below \SI{12}{\mega p\per\second}.
When exceeding a packet rate of around \SI{12}{\mega p\per\second}, all data streams started to lose packets, limiting the measured packet rate at that level as shown in Figure~\ref{fig:PacketRate2}.
Since the observed total maximum packet rate in this measurement is different from the one in the first measurement, a dependence on the packet size is inferred.

\subsection*{Measurement: influence of the packet size on the packet rate}

To investigate the dependence of the total maximum packet rate on the packet size, a third series of measurements was done.
Compared to the second series of measurements, the packet sizes was varied.
The packet size was chosen to be a power of two up to \SI{2048}{\byte}\footnote{\SI{2048}{\byte} is the largest packet size being a power of two and being supported by XDP.} and multiples of \SI{64}{\byte} up to \SI{512}{\byte}.
In addition, packet sizes one Byte larger than the aforementioned values are tested.

\begin{figure}[t]
  \centering
  \includegraphics[width=1.0\linewidth]{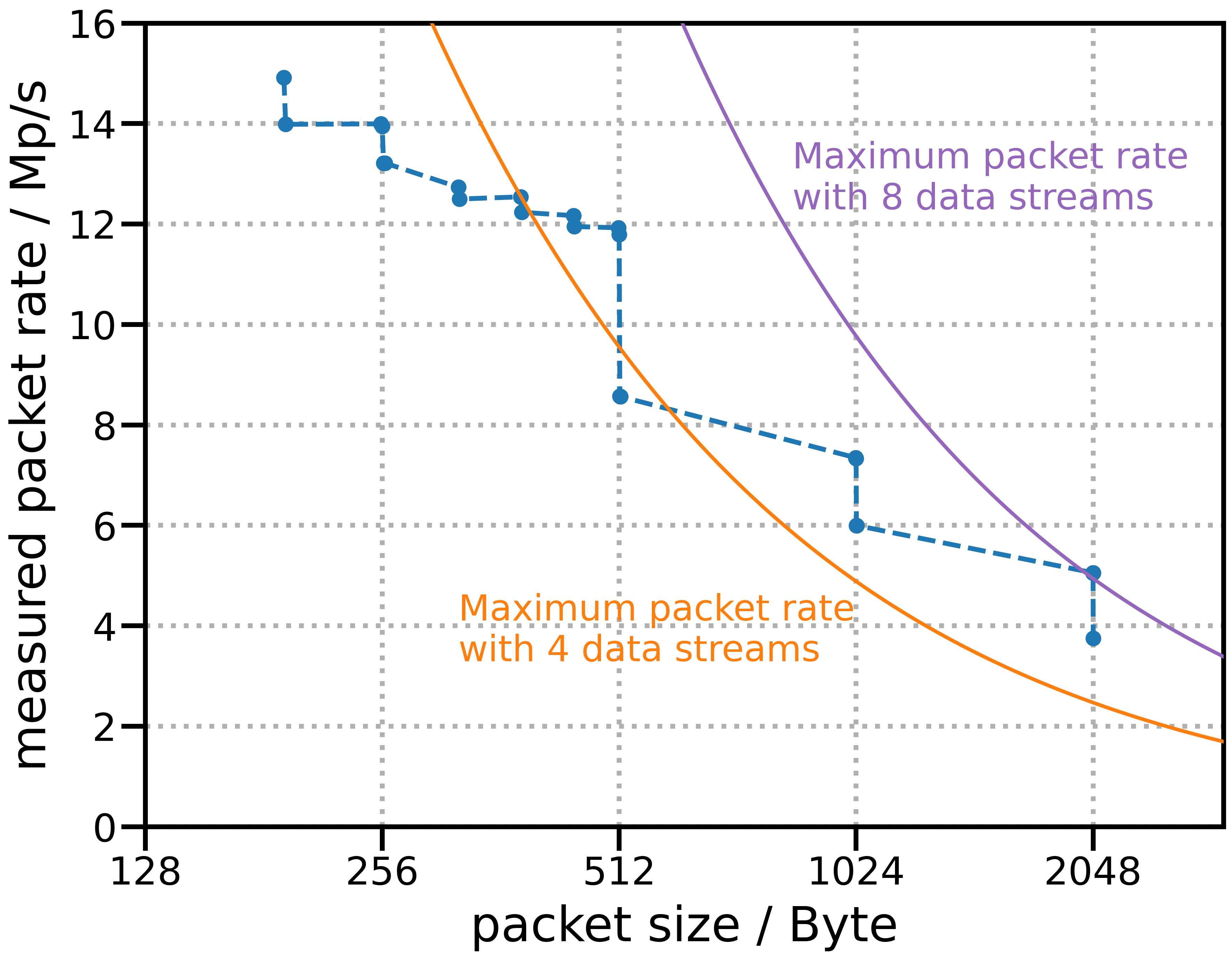}
  \caption{The measured total packet rates for different payload sizes. The measurements revealed steps in the achievable packet rate at packet sizes being a multiple of \SI{64}{\byte} up to \SI{512}{\byte} as well as at \SI{1024}{\byte} and \SI{2048}{\byte}. The packet sizes below \SI{192}{\byte} are not shown as their results were at the same level as for \SI{192}{\byte}. In addition to this, the maximum total packet rates for four and eight data generators are included. Only a small range of packet sizes is able to reach the full data rate of eight data generators.}
  \label{fig:PacketRate3}
\end{figure}

The measurements of the third series revealed several steps in the achievable packet rates as shown in Figure~\ref{fig:PacketRate3}.
The steps were found for each aforementioned pair of packet sizes.
For example, the server can handle packets of \SI{257}{\byte} at a notably lower rate than packets of size of \SI{256}{\byte}.
However, the reduction of the packet rate for each step shows no clear pattern.
In addition to the measured packet rates, the maximum packet rates that can be generated with four and eight data generators (i.e. around \SI{40}{\giga\bit\per\second} and \SI{80}{\giga\bit\per\second} respectively) are depicted.

\subsection*{Discussion}

An important result of the tests is the very narrow range of parameters to be used to achieve the optimal performance.
Only packet sizes just below \SI{2048}{\byte} were able to support the full payload data rate of around \SI{80}{\giga\bit\per\second} of the eight data generators used\footnote{A payload data rate of around \SI{80}{\giga\bit\per\second} is regarded as maximum reasonable data rate to leave some headroom for other network traffic.}, resulting in a packet rate of around \SI{5}{\mega p\per\second}.
For all other packet sizes, the packet rate had to be reduced to prevent packet loss.
This might restrict the possible patterns used to transfer data from the detector towards the processing system.
With most detector elements having no or only a very small number of hits to report, it can take some time to fill a complete frame and therefore might hit the timing requirements.
In such cases, it would be better to combine the responses of the same event from different detector elements instead of the event fragments of the same detector element from multiple events.

\section{Measurement: Optimization of task assignment}
\label{sec:optimization}
With the reachable packet rate of the server (i.e. around \SI{5}{\mega p\per\second}) being known from the previous set of measurements and the corresponding packet size (i.e. \SI{2048}{\byte}) to reach the intended overall data rate (i.e. a payload data rate of about \SI{80}{\giga\bit\per\second}), the question arises how to distribute the processing tasks across the CPU cores.
To minimize the number of CCX needed for the reception and decoding, the maximum data rate a sole CCX can handle is determined.
Thereby, the basic test setup stayed the same as used for the packet rate measurements.

The achievable performance is expected to depend on the assignment of the different threads to the available logical cores of a CCX.
Therefore, the possible combinations of thread assignments are evaluated and rated by the fraction of lost packets for different number of data streams (starting with a single one).
The goal is to find the combination with the highest data rate without losing packets.

After all measurements for a given number of data streams were conducted, some general rules for conflicting combinations (i.e. the combinations with packet loss) were derived.
These rules were then applied for the next round of measurements to keep the number of measurements at a manageable level.
If no combination for a given number of data streams was able to achieve a lossless data transmission, the combinations with the lowest packet loss were repeated with reduced data rate until no packets were lost or the overall data rate was below previously obtained rates.

In contrast to the previous measurements, the requirement to minimize the packet loss is to be taken into account for this and all the following measurements.
The payload size is set to \SI{2000}{\byte} as a result of the previous tests.
The data rate per stream is set to \SI{10}{\giga\bit\per\second} if no other value is specified.
The duration of each test run is limited to \SI{1}{\hour} for broader searches and to a few hours for some special cases.

Since the interrupt thread is fixed in position by the system configuration, only the assignment of the receiving thread and the worker threads for each data stream can be changed.
The four different positions are defined as:
\begin{itemize}
	\item \emph{Position 0} is the logical core on which the interrupt thread runs.
	\item \emph{Position 1} is the other logical core of the same physical core on which the interrupt thread is running on.
	\item \emph{Position 2} and \emph{position 3} are the logical cores of the second physical core of the CCX.
\end{itemize}
If multiple data streams are used, it is assumed that all worker threads behave the same and therefore the order of these can be ignored.
These threads are assigned with increasing stream number (i.e. \emph{stream 0} always gets a smaller position number than \emph{stream 1}).

\subsection*{Measurement: One stream}

For measurements of a single data stream per CCX, 16 combinations were tested since both the receiving thread and the worker thread were allowed on all four positions.
Since the CPU used contains eight CCX, eight of the combinations were tested in parallel.

With a single data stream, all combinations with the worker thread having its own position (i.e. not shared with other threads) showed no data loss.
As soon as the same position was shared with another thread (either the receiving or the interrupt thread), packet loss was seen.
Having the receiving and interrupt threads share the same position (i.e. position 0) did not generate packet loss.
The combinations with packet loss did not show a general pattern.
For example, no indication was found that the packet loss is smaller if the worker thread runs on the opposite physical core than the receiving thread.
Therefore, the only rule derived from these tests is that the worker threads need their own position which should not be shared with any other workload.

\subsection*{Measurement: Two streams}

When following the rule that a worker thread should not share its position with any other workload, the number of possible combinations
for the two-data-stream scenario is reduced drastically.
An additional feature of the setup is, that the ordering of the data streams and their associated worker threads is not changed, that is, the lowest stream number always runs on the lowest available position. Thus, only six combinations remain:
\begin{itemize}
	\item When the receiving thread runs on position 0, the two worker threads can run on any of the other three positions, resulting in three possible combinations (i.e. one combination for each unused position).
  \item When the receiving thread runs on any position other than 0, the worker threads are fixed in their position since they are not allowed to run on position 0, which always hosts the interrupt thread. This results again in three possible combinations.
\end{itemize}

In the case of two data streams, all tested combinations resulted in packets being lost.
The two combinations with both worker threads running on the second physical core (i.e. position 2 and 3) and the receiving thread running on position 0 or 1 had the highest fraction of lost packets (larger than \SI{30}{\percent}).
Both combinations with the first worker thread on position 1 and the second worker thread as well as receiving thread on the second physical core (i.e. each thread with its own position and both worker threads running on different physical cores) had the lowest packet loss observed at the level of a few \SI{10}{ppm}.

\subsection*{Measurement: Two streams, reduced data rates}

Since the two cases with the lowest packet loss had roughly the same fraction of lost packets, both were tested again with reduced data rates.
For both combinations, the data rates for both data streams were reduced alternately until no lost packets were observed over a period of \SI{12}{\hour}.

The best result was achieved with the receiving thread running on position 2 and the two worker threads running on position 1 and 3.
The total data rate reached \SI{15.5}{\giga\bit\per\second} (\SI{7.5}{\giga\bit\per\second} for the first stream and \SI{8.0}{\giga\bit\per\second} for the second stream).
As soon as the data rate of either stream was slightly raised (i.e. by \SI{0.5}{\giga\bit\per\second}) the first stream started to lose packets while the second stream did not.

The same was true when a third data stream was added to the tests, independent of its position.
As a result, the maximum achievable data rate per CCX is found to be \SI{15.5}{\giga\bit\per\second}.

\subsection*{Discussion}

The presented results confirm the use of a complete CPU for the data reception and processing.
With the histogramming of the data being only a very simple reflection of the real processing to be done, some additional processing capabilities should be reserved for the more complex data formats.
Therefore, it would be better to rather process only a single data stream on a CCX than to use this processing power for the reception of a second data stream.
This helps in keeping the data locally in the cache.
In addition to this, the real setup would require the transmission of the data to the next processing step which will also occupy processing resources and was not done in the test setup presented.

\section{Measurement: Long duration tests}
In this set of measurements, the duration of the measurements was increased to include spurious effects that may occur only very rarely.

\subsection*{Measurement}

For the long duration tests eight data streams with the maximal data rate of \SI{10,24}{\giga\bit\per\second} each were used, resulting in a payload data rate of around \SI{80}{\giga\bit\per\second}, not including any protocol overhead.
Each data stream was assigned to a separate CCX since the measurements at maximum data rate showed that a single CCX cannot handle two of these streams without losing packets.

The data generators produced a \SI{16}{\bit} counting pattern and the processing was implemented by counting the occurrence of each \SI{16}{\bit} word.
Since the block size is a multiple of \SI{128}{\kilo\bit}, each counter value has the same frequency in the data stream.
The used payload size per packet of \SI{2000}{\byte}, however, is not equal to a power of two (i.e. only having a relative small common factor with respect to the block size) and thus the time period is increased until the identical payload is seen again.

\begin{figure}[!tb]
  \centering
  \includegraphics[width=0.9\linewidth]{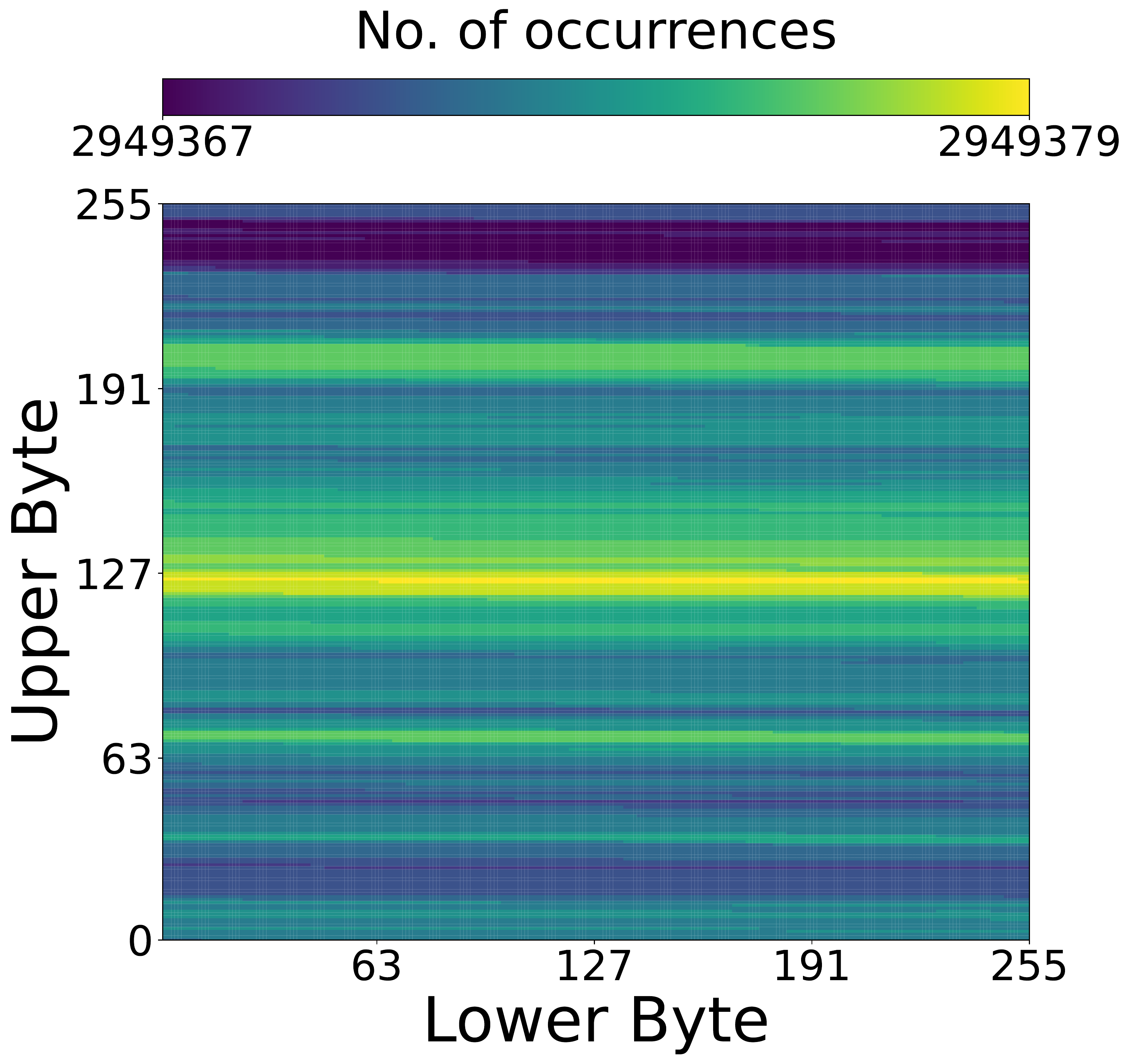}
  \caption{An example of a histogram of transmitted data words for a flawed data transmission.
    The histogram counts the occurrence of \SI{16}{\bit} words received in the payload.
    The words are split into an upper and a lower byte both of which are used as indices defining a two-dimensional array.
    The strips of different (darker) color indicate the loss of data words. 
    A lossless transmission, in contrast, would result in two monochromatic areas with the counter values for both
    areas being off by one.
    In this example, bit flips in the payload resulted in invalid checksums and complete packets were dropped.
    Since a payload size of \SI{2000}{\byte} was used, a packet corresponds to slightly less than four lines
    in the histogram.}
  \label{fig:Histo}
\end{figure}

Histogramming of the data words was used as a method to check the completeness and correctness of the transferred data.
The \SI{16}{\bit} words are split into an upper and a lower byte.
These two bytes are used as indices in a two-dimensional array with the entry at the given position holding the number of occurrences of the corresponding \SI{16}{\bit} word.
When the histogram is plotted, each data word is represented by a pixel with a color corresponding to the value of its counter.
In case of a lossless data transmission, the difference between the highest and lowest occurrence of data words seen is one at most.
The difference occurs for the cases in which the amount of transferred data is not exactly a multiple of \SI{128}{\kilo\bit} since the test is aborted at a random point\footnote{The data generators were configured to start the counting pattern with a value of 0 and were started after the receiving programs were ready. Therefore, there cannot be a second step resulting from a misaligned starting point.}.
Therefore, the plotted histogram for a lossless data transmission features at most two homogenously filled areas.
In contrast to a successful test, an example histogram of a flawed data transmission is depicted in Figure~\ref{fig:Histo}.
In this test, data words were lost at the level of \(10^{-6}\).
The histogram shows a clear indication of unevenly distributed occurrences of data words.

The duration of the long-term measurements was set to seven days.
If a notable packet loss occurred, the measurement was aborted to investigate the cause of it.
This was repeated until an acceptable fraction of packet loss was achieved.

\subsection*{Results}

The final long duration test ran seven days with a total amount of \SI{5.2}{\peta\byte} of data transferred within 2.92 trillion packets.
During this test, not a single packet was lost in any of the eight data streams.
To give an upper limit on the ratio of lost to transferred data packets, it is assumed that the next packet got lost, resulting in a ratio of \num{3.423e-13}.
This result was reproduced in tests with a similar duration.

The correctness of the result was checked by multiple methods.
First of all, the reporting of the receiving software was checked by provoking an error situation.
An instance of the receiving software was started and kept running until the end of the test.
Then, a data stream was started, sent towards the previously started software instance and stopped at a random point.
Afterwards, a second data stream with a different value of its packet identifier field was started and sent towards to same software instance.
The result of this test was the correct detection of the jump of the value of the packet identifier field.

The plotted histograms generated by the worker threads can also be used to check the correctness of the result as described in the measurements section.
In case of the final long duration test, all these plots were divided only into two monochromatic areas with the counter values for both being off by one\footnote{Each stream had its own set of values since the streams were not started and stopped exactly simultaneously and therefore the amount of transferred data differs from stream to stream.}.
This does not only prove that no packet was lost in between, but also that no bit got flipped\footnote{The occurrence of only combinations that cancel each other out is unlikely and therefore ignored.}.

Another way to verify the result is to sum up all counter values of a histogram.
Since a data word contains two bytes, the sum should thus be equal to half the number of bytes transferred by the corresponding data stream.
This third method also confirmed the result of no packet being dropped. The amount of transferred data per data stream was around \SI{700}{\tera\byte}.

\subsection*{Discussion}

The most important result of the performed tests is that not a single packet was dropped during the long duration tests.
This is the optimal outcome and was achieved by using the XDP feature of the Linux kernel, leading to a significant reduction of the processing overhead.
Thus, the proposed system allows for a highly reliable data transfer, while the usage of protocols with guaranteed data transmission like TCP offer little additional benefit, but requires much larger system resources.
The presented approach based on direct data transmission from the FPGA to the DAQ servers via a UDP-based protocol offers a valid opportunity to reduce the overall system complexity and system costs.

\section{Conclusion}
In this paper, a novel approach of transferring data between FPGAs and commodity servers was studied.
This alternative is based on waiving guaranteed data transmission to enable a reduction of the system complexity and system costs.
The XDP technique was used for the first time in the context of high energy physics to accelerate the data reception in the DAQ systems.
XDP helps to reduce the CPU workload and therefore plays a key role in achieving a lossless transmission of \SI{5.2}{\peta\byte} of data with a payload data rate of \SI{80.0}{\giga\bit\per\second}.
This result proves the validity of this alternative approach.

The studies showed that the performance of the data transmission in terms of the achieved payload rate 
depends critically on the chosen packet size.
The maximum packet rate is a feature of a given server, providing a certain number of CPU cores.
The rate depends on the tasks of the CPU cores and has to be determined for a given configuration.
Discontinuous strong drops in the packet rate are observed when passing powers of \num{2} in the packet size.
To obtain an optimized transmission for a specific server configuration, the packet rate and the packet size
need to be carefully adjusted.
For the specific setup under study optimal performance is observed for a packet size just
below \SI{2048}{\byte}, the largest packet size supported by XDP.

Furthermore, the system performance depends critically on the assignment of tasks to the logical cores of the CCX.
Processing a single data stream per CCX is recommendable, even though the processing of two streams in a single
CCX is found to be possible without losses if the data rates are reduced accordingly.

The next step of further evaluating the proposed system will be the use of realistic data patterns as expected from the tracking detectors at the LHC.
This includes the use of the real data format being generated by the detector elements and the corresponding decoding routines.
Since the parameters for the data transmission had to be optimized for the system used, the optimization process needs to be repeated as soon as the system is changed, for example if a different type of CPU is used.

\section*{Acknowledgments}
This work was supported by the German Federal Ministry of Education and Research.
We also wish to thank Gerhard Brandt, Frank Ellinghaus and Dominic Hirschbuehl for useful discussion.
\printbibliography

\end{document}